# Design, Modelling, and Control of Magnetic Ball Suspension System

Sampson E. Nwachukwu, *Member, IEEE*

*Abstract*—This paper presents the modeling, control design, and performance analysis of a Magnetic Ball Suspension System (MBSS), a nonlinear and inherently unstable electromechanical system used in various precision applications. The system's primary objective is to levitate a steel ball using electromagnetic force without physical contact, thereby eliminating frictional losses. A comprehensive state-space model was developed, capturing both the mechanical and electrical dynamics. The equilibrium points of the system were determined through feedback linearization using the Jacobian matrix. To ensure system stability, controllability and observability analyses were conducted, confirming that state feedback and observer-based control strategies could be effectively implemented. Three distinct control methods were explored: pole placement-based state feedback control, full-order observer design, and optimal state feedback control using the Linear Quadratic Regulator (LQR). Each control strategy was validated through Simulink simulations for both linearized and nonlinear models. Simulation results demonstrated that the linearized system consistently achieved desired performance with minimal oscillations, whereas the nonlinear system exhibited significant transient oscillations before stabilization. The full-order observer enhanced estimation accuracy, enabling effective control where direct state measurement was impractical. The LQR-based control offered improved robustness and minimized control effort, though its performance was comparable to standard state feedback in some cases.

## I. Introduction

Electro-mechanical systems that allow an item to float in a certain area without the need for support are called magnetic ball suspension systems (MBSS). This technology is widely used in wind turbines, micro-robot actuation, and very precise positioning systems due to its friction-free and contact-free qualities, which eliminate the loss of energy brought on by friction [1]. However, the design, modeling, and analysis of the behavior and operating stability of the MBSS are complicated due to the unstable nature and innate nonlinearities of the systems, caused by the volatility of the input signals that are always changing. As a result, the MBSS needs adequate control action. Hence, the majority of the analysis effort has been done on the controller design [1], [2].

Over the years, a number of dynamic models of magnetic force of MBSS have been proposed. With these models, different linear and nonlinear control techniques have been applied, and their respective performances have been compared. The linear system concept is only effective within limited operating point ranges [3], [4].

Because the MBSS is nonlinear in nature, different criteria must be considered in its design. Therefore, the goal of this study is to design the MBSS and analyze its behavior and stability under different operating scenarios. The design methodology involves developing the state space model of the system and determining its equilibrium points using the feedback linearization, with the aim of controlling the ball's location to ensure that the system performs satisfactorily. With the state-space model approach, the system can be analyzed in the time domain by using a set of first-order differential equations. This eliminates the need to do partial fraction calculations and the complex Laplace transforms in order to analyze a control system. The state space model is a mathematical representation of a system that consists of a set of inputs, outputs, and state variables coupled by a first-order differential equation [1], [2].

In addition, this study covers the observability and controllability of the system, assuming that the system can be observed. If every state of a system can be moved to any location with an input, it is said to be controllable. On the other hand, if every state of a system can be observed, meaning that its values can be ascertained from the output, it is said to be observable. Thus, controllability guarantees that the poles or eigenvalues of a system may be moved to any required location in order to achieve the stability or optimality of the system. This can be done through state feedback. The state feedback is made possible by observability, which guarantees that we can estimate or reconstruct the state variables from the output [5].

However, it is not possible to measure every state precisely in practice. Thus, the state feedback cannot be used directly in most applications. Therefore, this study delved into developing a full-order and optimal observers which allows us to "estimate" the variables of the state and the application of feedback derived from estimations of the state. This can only be achieved when the system is observable [5]. Furthermore, the optimal state feedback is presented based on linear quadratic regulator (LQR), which is capable of handling wide range of applications.

The remainder of this study is organized as follows: Section II presents the mathematical formulation of MBSS. Section III presents the system control design methodologies, Section IV presents the simulation results. Finally, Section IV presents the conclusion of the study.

## II. Mathematical Formulation of MBSS

### a. Dynmanic Modelling of MBSS

In the phase plane, an equilibrium point is called a singular point. This suggests that $\dot{x} = 0$ as an equilibrium point is defined as a location where the system states may remain indefinitely. The equilibrium states can be determined using the following equations [3].

$$\dot{x}_1 = f_1(x_1, x_2 ..., x_n) \quad (1)$$
$$\dot{x}_2 = f_2(x_1, x_2 ..., x_n) \quad (2)$$

While a continuous series of single points may exist in some circumstances, a linear system typically has just one singular point [3]. As illustrated in this study, nonlinear systems, such as MBSS, frequently include several isolated single points, and it is crucial to determine these points to understand the operating stability of the system at all times. The schematic diagram of the MBSS is shown in Fig. 1. By varying the current $i$ in the electromagnet via the input voltage $e(t)$, the position, $y$ of the steel ball with mass, $M$ may be manipulated. A series Resistor-Inductor (RL) circuit with winding inductance $L$, and winding resistance, $R$ is used to illustrate the electromagnet. The force, $F$ generated by the electromagnet is represented by $\frac{Ki^2}{y^2}$, where $K$ is the constant proportional to the electromagnet's cross-section area and turn ratio. Controlling the electromagnet's current is the system's main goal. This helps to maintain the ball's fixed suspension length from the magnet's end [6].

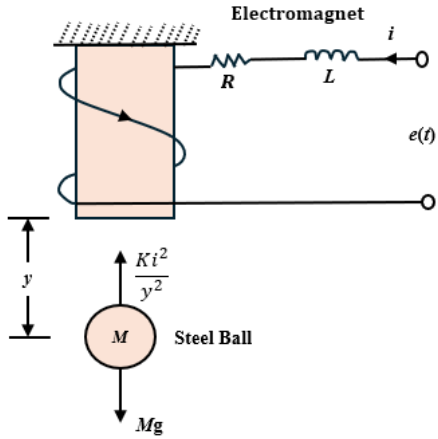

Fig. 1. Schematic diagram of MBSS

The dynamic modelling of the MBSS may be represented as the derivative of the equation of motion of the steel ball, which is generated by the motion of the steel ball, and the electric current's equation that makes up an electromagnet [1].

*a) Mechanical Part Dynamic Modelling*

Recall that $\dot{y}$ is the ball's velocity, and $F = M\ddot{y}$, where $\ddot{y}$ is the acceleration. Therefore, the dynamic model of the mechanical part of the system is determined by applying the following Newton's 2nd law:

$$M\ddot{y} = Mg - \frac{Ki^2}{y^2} \quad (3)$$

where g represents the gravitational acceleration.

*b) Electrical Part Dynamics Modelling*

Similarly, the dynamic model of the electrical part of the system is determined by applying the following Kirchhoff's Voltage Law (KVL):

$$e(t) = Ri + L\frac{di}{dt} \quad (4)$$

Since the system is nonlinear in nature, the two essential variables in (3) are the electromagnetic coil's current flow, $i$ and the distance, $y$ between the electromagnet and the ball. Thus, the total inductance is deremined as:

$$L(y) = \frac{L}{y} \quad (5)$$

From (5), it is clear that the inductor, $L$ is depending on $y$, and it is not constant.

*c) State Space Model*

For the electrical systems, the voltages across the capacitors and the current across the inductors are chosen as the state variables. In contrast, for mechanical systems, the positions and velocities of the rigid bodies are chosen as the state variables. Thus, the differential equation model of the MBSS is determined by considering the state variables and inputs. Subsitituting (5) into (4), the differential equation model of the electrical part of the system is derived as:

$$\frac{di}{dt} = \frac{y}{L}(e(t) - Ri) \quad (6)$$

Also, the differential equation model of the mechanical part of the system is derived as:

$$\ddot{y} = g - \frac{Ki^2}{My^2} \quad (7)$$

Thus, the state variables are:

$$\begin{cases} x_1 = y \\ \dot{x}_1 = \dot{y} = x_2 = f_1 \\ \dot{x}_2 = \ddot{y} = g - \frac{Ki^2}{My^2} = f_2 \\ \dot{x}_3 = \frac{di}{dt} = \frac{y}{L}(e(t) - Ri) = f_3 \end{cases} \quad (8)$$

Thus, the state-space model is:

$$\dot{x} = \begin{bmatrix} \dot{x}_1 \\ \dot{x}_2 \\ \dot{x}_3 \end{bmatrix} = \begin{bmatrix} x_2 \\ g - \frac{Kx_3^2}{Mx_1^2} \\ \frac{x_1}{L}(e(t) - Rx_3) \end{bmatrix} \quad (9)$$

$$y = [x_1 \ x_2 \ x_3]^T = [1 \ 0 \ 0] \quad (10)$$

*d) Equilibrium Points of MBSS*

The state variables, $y$ and $\dot{y}$ are constant over a time, since, at equilibrium, the system is stable. Thus, we set the state's derivatives to zero in order to determine the nominal value of $y$ and $\dot{y}$ at equilibrium. Also, we set $E$ to be the nominal value of $e(t) = u$, and $\begin{bmatrix} x_{10} \\ x_{20} \\ x_{30} \end{bmatrix} = \begin{bmatrix} y_0 \\ \dot{y}_0 \\ i_0 \end{bmatrix}$. Therefore, around the equilibrium points,

$$\dot{x}_1 = \dot{y} = x_{20} = 0 \quad (11)$$
$$\dot{x}_2 = g - \frac{Kx_{30}^2}{Mx_{10}^2} = 0 \quad (12)$$
$$\dot{x}_3 = \frac{x_{10}}{L}(E - Rx_{30}) = 0 \quad (13)$$

Since $R$ is the electromagnet's direct resistance and the available nominal voltage to supply an electric current, $i_0$ is $E$, their relationship is $i_0 = x_{30} = \frac{E}{R}$. Thus, $x_{10}$ in (12) becomes:

$$x_{10} = \frac{E}{R}\sqrt{\frac{K}{Mg}} \tag{14}$$

Thus, $\frac{dy}{dt} = \frac{dx_{10}}{dt} = 0$, and the equilibrium points of the system are:

$$\begin{bmatrix} x_{10} \\ x_{20} \\ x_{30} \end{bmatrix} = \begin{bmatrix} \frac{E}{R}\sqrt{\frac{K}{Mg}} \\ 0 \\ \frac{E}{R} \end{bmatrix} \tag{15}$$

In this study, we set $E = 8V$, $R = 10$ ohms, $K = 0.01$, $M = 0.2$ Kg, $L = 0.5$ H, and $g = 9.8$. Using these values, the equilibrium points are calculated as:

$$\begin{bmatrix} x_{10} \\ x_{20} \\ x_{30} \end{bmatrix} = \begin{bmatrix} 0.06 \\ 0 \\ 0.8 \end{bmatrix} \tag{16}$$

e) *Linearization of MBSS*

The nonlinear system's stability can be determined by using the feedback linearization approach or the Lyapunov functions approach. In this study, we considered the feedback linearization approach based on the Jacobian Matrix. This is due to the Jacobian Matrix's suitability for the observer design, which will be discussed in detail later in this study.

Defining the derivation of the system from the nominal operating point, we have:

$$\begin{cases} \Delta x_1 = x_1 - x_{10} = x_1 - \frac{E}{R}\sqrt{\frac{K}{Mg}} \\ \Delta x_2 = x_2 - x_{20} = x_2 \\ \Delta x_3 = x_3 - x_{30} = 0 \\ \Delta u = u - E = 0 \end{cases} \tag{17}$$

Also, the goal is to linearize the system around the equilibrium points using the following Jacobian matrix:

$$\begin{bmatrix} \Delta\dot{x}_1 \\ \Delta\dot{x}_2 \\ \Delta\dot{x}_3 \end{bmatrix} = \begin{bmatrix} \frac{\partial f_1}{\partial x_1} & \frac{\partial f_1}{\partial x_2} & \frac{\partial f_1}{\partial x_3} \\ \frac{\partial f_2}{\partial x_1} & \frac{\partial f_2}{\partial x_2} & \frac{\partial f_2}{\partial x_3} \\ \frac{\partial f_3}{\partial x_1} & \frac{\partial f_3}{\partial x_2} & \frac{\partial f_3}{\partial x_3} \end{bmatrix} \begin{bmatrix} \Delta x_1 \\ \Delta x_2 \\ \Delta x_3 \end{bmatrix} + \begin{bmatrix} \frac{\partial f_1}{\partial u} \\ \frac{\partial f_2}{\partial u} \\ \frac{\partial f_3}{\partial u} \end{bmatrix} \begin{bmatrix} \Delta u \\ \Delta u \\ \Delta u \end{bmatrix} \tag{18}$$

$$\Delta\dot{x} = A^*\Delta x + B^*\Delta u$$

$$\Delta\dot{y} = C^*\Delta x + D^*\Delta u$$

where $\Delta x$ represents the state vector, and $\Delta u$ is the linearized system's input voltage.

Thus, using the Jacobian matrix in (18), the system in (8) is linearized around its equilibrium pints as follows:

$$\Delta\dot{x}_1 = \frac{\partial f_1}{\partial x_1}(x_{10}, x_{20}, x_{30})\Delta x_1 + \frac{\partial f_1}{\partial x_2}(x_{10}, x_{20}, x_{30})\Delta x_2 + \frac{\partial f_1}{\partial x_3}(x_{10}, x_{20}, x_{30})\Delta x_3 + \frac{\partial f_1}{\partial u}(x_{10}, x_{20}, x_{30})\Delta u$$

$$= 0 + \Delta x_2 + 0 + 0 = \Delta x_2$$

$$\Delta\dot{x}_2 = \frac{\partial f_2}{\partial x_1}(x_{10}, x_{20}, x_{30})\Delta x_1 + \frac{\partial f_2}{\partial x_2}(x_{10}, x_{20}, x_{30})\Delta x_2 + \frac{\partial f_2}{\partial x_3}(x_{10}, x_{20}, x_{30})\Delta x_3 + \frac{\partial f_2}{\partial u}(x_{10}, x_{20}, x_{30})\Delta u$$

$$= \frac{2Kx_{30}^2}{MKx_{10}^3}\Delta x_1 + 0 - \frac{2Kx_{30}}{MKx_{10}^2}\Delta x_3 + 0$$

$$= \frac{2Kx_{30}^2}{MKx_{10}^3}\Delta x_1 - \frac{2Kx_{30}}{MKx_{10}^2}\Delta x_3 = 296.29\Delta x_1 - 22.2\Delta x_3$$

$$\Delta\dot{x}_3 = \frac{\partial f_3}{\partial x_1}(x_{10}, x_{20}, x_{30})\Delta x_1 + \frac{\partial f_3}{\partial x_2}(x_{10}, x_{20}, x_{30})\Delta x_2 + \frac{\partial f_3}{\partial x_3}(x_{10}, x_{20}, x_{30})\Delta x_3 + \frac{\partial f_3}{\partial u}(x_{10}, x_{20}, x_{30})\Delta u$$

$$= \frac{1}{L}(E - E)\Delta x_1 + 0 - \frac{Rx_{10}}{L}\Delta x_3 + \frac{x_{10}}{L}\Delta u$$

$$= -\frac{Rx_{10}}{L}\Delta x_3 + \frac{x_{10}}{L}\Delta u = -1.2\Delta x_3 + 0.12\Delta u$$

From the above, the linearized system is:

$$\begin{bmatrix} \Delta\dot{x}_1 \\ \Delta\dot{x}_2 \\ \Delta\dot{x}_3 \end{bmatrix} = \begin{bmatrix} 0 & 1 & 0 \\ \frac{2Kx_{30}^2}{MKx_{10}^3} & 0 & \frac{2Kx_{30}}{MKx_{10}^2} \\ 0 & 0 & \frac{Rx_{10}}{L} \end{bmatrix} \begin{bmatrix} \Delta x_1 \\ \Delta x_2 \\ \Delta x_3 \end{bmatrix} + \begin{bmatrix} 0 \\ 0 \\ \frac{x_{10}}{L} \end{bmatrix} \begin{bmatrix} \Delta u \\ \Delta u \\ \Delta u \end{bmatrix} \tag{19}$$

Substituting the calculated values of $x_{10}$, $x_{20}$, and $x_{30}$ in (16), the linearized system is expressed as:

$$\begin{bmatrix} \Delta\dot{x}_1 \\ \Delta\dot{x}_2 \\ \Delta\dot{x}_3 \end{bmatrix} = \begin{bmatrix} 0 & 1 & 0 \\ 296.29 & 0 & -22.2 \\ 0 & 0 & -1.2 \end{bmatrix} \begin{bmatrix} \Delta x_1 \\ \Delta x_2 \\ \Delta x_3 \end{bmatrix} + \begin{bmatrix} 0 \\ 0 \\ 0.12 \end{bmatrix} \begin{bmatrix} \Delta u \\ \Delta u \\ \Delta u \end{bmatrix} \tag{20}$$

Also, the output of the system is expressed as:

$$\Delta\dot{y} = [1\ 0\ 0]\begin{bmatrix} \Delta x_1 \\ \Delta x_2 \\ \Delta x_3 \end{bmatrix} \tag{21}$$

From (20),

$$A^* = \begin{bmatrix} 0 & 1 & 0 \\ 296.29 & 0 & -22.2 \\ 0 & 0 & -1.2 \end{bmatrix}; B^* = \begin{bmatrix} 0 \\ 0 \\ 0.12 \end{bmatrix};$$

$$C^* = [1\ 0\ 0]; D^* = 0$$

b. *Controllability and Observability of MBSS*

1. *Controllability of MBSS*

The controllability matrix of the system is expressed as:

$$C = [B^*\ A^*B^*\ A^{*2}B^*\ \ldots\ A^{*\,n-1}B^*] \tag{22}$$

For a system to be controllable, the system's controllability matrix must have a full rank. That is,

$$(A^*, B^*) \text{ is controllable} \Leftrightarrow \text{rank}(C) = n$$

Thus, the controllability matrix of the system is calculated as:

$$C = \begin{bmatrix} 0 & 0 & -2.6640 \\ 0 & -2.6640 & 3.1968 \\ 0.1200 & -0.1440 & 0.1728 \end{bmatrix}$$

The system is controllable since the rank of $C$ is 3.

2. *Observability of MBSS*

The observability matrix of the system is expressed as:

$$O = \begin{bmatrix} C^* \\ C^*A^* \\ \vdots \\ C^{*n-1}A^* \end{bmatrix}; \quad (23)$$

For a system to be observable, the system's observability matrix must also have a full rank. That is,

$(A^*, C^*)$ is controllable $\Leftrightarrow$ rank$(O) = n$

The observability matrix of the system is calculated as:

$$O = \begin{bmatrix} 1 & 0 & 0 \\ 0 & 1 & 0 \\ 296.29 & 0 & -22.20 \end{bmatrix}$$

The system is observable since the rank of $O$ is 3.

### III. SYSTEM CONTROL DESIGN

#### a. State Feedback Control of MBSS

The placement of a system's poles or eigenvalues has a direct impact on its stability and optimality. The question that mostly arise is whether it possible to reposition the poles of the system if they are not where we would like them to be? Pole placement approach is used in this study to tackle this particular challenge. By using feedback control, pole placement may be accomplished [7].

The eigenvalues of $A^*$, represented by $\lambda(A^*)$, are the poles of the system. Here, we employ the state feedback control formula, $u = Kx + v$, in which $v$ represents an external input and $Kx$ is the linear state feedback. The controlled system is determined by this feedback control expressed as [7]:

$$\dot{x} = (A^* + B^*K) + B^*v \quad (24)$$

where $(A^* + B^*K)$ are the poles of the controlled system.

Since $(A^*, B^*)$ is controllable, based on our calculations earlier, the poles of the system in the complex plane can be moved to any desired location using the following steps [7]:

i. Determine the characteristics polynomial, $\varphi(s)$ of $A^*$.

$$\varphi(s) = (SI-A^*) = s^n + a_{n-1}s^{n-1} + \cdots + a_1 s + a_0 \quad (25)$$

where $a_0, a_1$, are the coefficients of the characteristic polynomial. Thus,

$$\varphi(s) = s^3 + 1.2s^2 - 296.29s - 355.548$$

ii. Present the state space representation of the system in the controllable canonical form $(A^*_c, B^*_c)$:

$$\dot{x} = \begin{bmatrix} 0 & 1 & \cdots & 0 \\ \vdots & \vdots & \cdots & \vdots \\ 0 & 0 & \cdots & 1 \\ -a_0 & -a_1 & \cdots & -a_{n-1} \end{bmatrix} x + \begin{bmatrix} 0 \\ \vdots \\ 0 \\ 1 \end{bmatrix} u \quad (26)$$

$$= \begin{bmatrix} 0 & 1 & 0 \\ 0 & 0 & 1 \\ 355.548 & 296.29 & -1.2 \end{bmatrix} x + \begin{bmatrix} 0 \\ 0 \\ 1 \end{bmatrix} u$$

where, $A^*_c = \begin{bmatrix} 0 & 1 & 0 \\ 0 & 0 & 1 \\ 355.548 & 296.29 & -1.2 \end{bmatrix}$, and

$$B^*_c = \begin{bmatrix} 0 \\ 0 \\ 1 \end{bmatrix}$$

iii. Determine the desired characteristics polynomial, $\bar{\varphi}(s)$ of the system:

$$\bar{\varphi}(s) = s^n + \bar{a}_{n-1}s^{n-1} + \cdots + \bar{a}_1 s + \bar{a}_0 \quad (27)$$

Placing the pole of the system at -5, -10, -20,

$$\bar{\varphi}(s) = (s+5)(s+10)(s+20) = s^3 + 35s^2 - 350s - 1000$$

iv. Determine the transform matrix, $T_c$:

$$T_c = C * \left[[B^*_c \; A^*_c B^*_c \; A^{*2}_c B^*_c \; \ldots \; A^{*n-1}_c B^*_c]\right]^{-1} \quad (28)$$

$$= \begin{bmatrix} -2.6640 & 0 & 0 \\ 0 & -2.664 & 0 \\ -35.5548 & 0 & 0.12 \end{bmatrix}$$

v. Determine the feedback matrix, $K_c$ for $(A^*_c, B^*_c)$:

$$K_c = [a_0 - \bar{a}_0 \; a_1 - \bar{a}_1 \; \ldots \; a_{n-1} - \bar{a}_{n-1}] \quad (29)$$

$$= [-1,355.5 \; -646.3 \; -33.8]$$

vi. Determine the feedback matrix, $K$ for $(A^*, B^*)$:

$$K = K_c T_c^{-1} \quad (30)$$

$$= [4,268.1 \; 242.6 \; -28.17]$$

vii. Develop the state feedback control:

$$u = Kx + v \quad (31)$$

The Simulink model of linearized and nonlinearized system state feedback control is shown in Fig. 2 and Fig. 3, respectively.

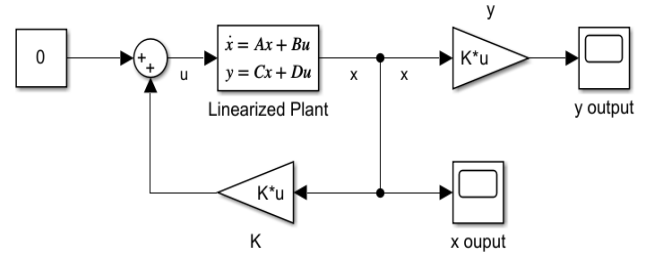

Fig. 2. Simulink model of linearized system state feedback control

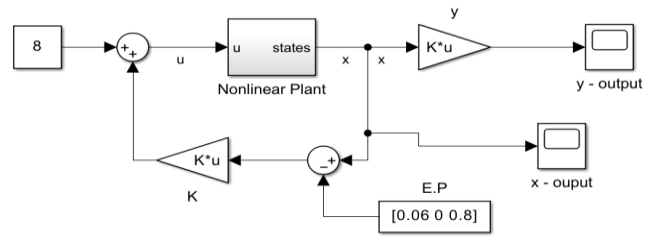

Fig. 3. Simulink model of a nonlinear system state feedback control

#### b. Full-Order Observer Design for MBSS Control

In order to attain stability and optimality, poles can be placed at any point in the complex plane using state feedback in a controlled system, as discussed earlier. However, it is not possible to measure every state exactly in practice. For this reason, state feedback cannot be used directly in most applications. Therefore, in this section, an observer is designed to estimate the system's state. This is only achievable if the system $(A^*, C^*)$ is observable. The observer's dynamics can be determined using the following equation [7]:

$$\dot{\hat{x}} = (A^* + GC^*)\hat{x} + B^*u - Gy \quad (32)$$

where $G$ is the observer matrix used to provide error feedback, with the following differential equation satisfying the estimating error, $\tilde{x}$.

$$\dot{\tilde{x}} = \dot{x} - \dot{\hat{x}} = (A^* + GC^*)x - (A^* + GC^*)\hat{x}$$
$$= (A^* + GC^*)\tilde{x} \quad (33)$$

In (33), the eigenvalues of $(A^* + GC^*)$ dictate the dynamics of $\tilde{x}$. Also, the observer's poles can be positioned appropriately to satisfy the required performance standards as the observer matrix $G$ can be chosen during the design process. Thus, it is possible to solve the observer's pole placement problem as a "dual" of the state feedback's pole placement problem. It should be noted that the eigenvalues of a matrix and its transpose are the same [7]. Thus,

$$\lambda(A^* + GC^*) = \lambda(A^{*T} + C^{*T}G^T) \quad (34)$$

Therefore, the observer's pole placement problem is equivalent to the pole placement problem for the state feedback if we consider $(A^{*T}, C^{*T})$ as $(A^*, B^*)$ and $G^T$ as $K$. This can be achieved only when $(A^*, B^*)$ is observable and $(A^{*T}, C^{*T})$ is controllable. By applying the duality approach, the full-order observer design can be designed using the following steps [7]:

i. Apply steps i – iii in subsection A.

ii. Determine the transform matrix, $T_c$:

$$T_c = O^T * \left[ [B^*_c \ A^*_c B^*_c \ A^{*2}B^* \ ... \ A^{*n-1}_c B^*_c] \right]^{-1} \quad (35)$$

iii. Apply steps e and f in subsection A.

iv. Determine the observer matrix, $G$.

$$G = K^T \quad (36)$$

v. Develop the observer:

$$\dot{\hat{x}} = (A^* + GC^*)\hat{x} + B^*u - Gy \quad (37)$$

The Simulink model of linearized and nonlinear system observer is shown in Fig. 4 and Fig. 5, respectively.

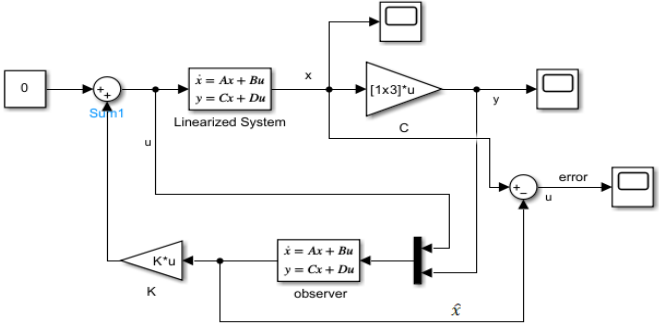

Fig. 4. Simulink model of feedback control of a linearized system using an observer

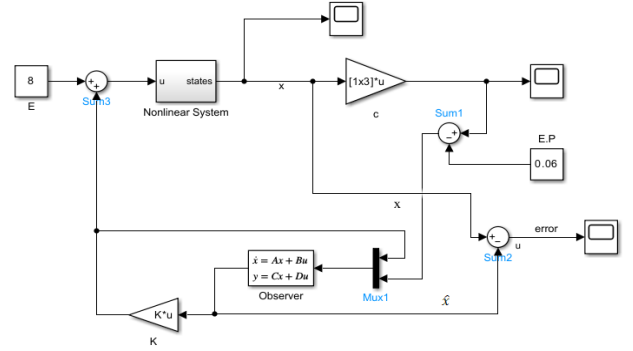

Fig. 5. Simulink model of feedback control of a nonlinear system using an observer

### c. Optimal State Feedback Control

Generally, the linear quadratic regulator (LQR) problem is dual to the optimal state feedback or observation problem. Here, developing a control that reduces the following "cost function" is the main objective [7].

$$J(x, t) = \int_t^{t_f} L(x, u) \, d\tau \quad (38)$$

where $L(x, u)$ describes the cost goal, $t$ is the current time, $t_f$ is the termination time, and $x = x(t)$ is the current state.

The aforementioned cost function is quite broad and is capable of handling a wide range of practical control problems. For a linear time-invariant system, i.e., $\dot{x} = A^*x + A^*u$, with desired state value $x_d = 0$, the cost function is quadratic and is determined as follows [7]:

$$J(x, t) = \int_t^{t_f} L(x^T Q x + u^T R u) \, d\tau \quad (39)$$

Then, an LQR problem is the term used to describe the optimal control problem. Here, $Q = Q^T \geq 0$ and $R = R^T \geq 0$ are "symmetric and positive semi-definite" matrix. The minimal cost is assumed to be quadratic in order to solve the LQR problem. Thus,

$$J^*(x, t) = x^T S(t) x \quad (40)$$

By applying the Hamilton–Jacobi–Bellman equation in [7], the optimal control is determined as:

$$u^* = -R^{-1} B^{*T} S(t) x \quad (41)$$

where $S(t)$ satisfies the following Riccati equation:

$$\dot{S}(t) = -(S(t) A^* + A^{*T} S(t) + Q - S(t) B^* R^{-1} B^{*T} S(t)) \quad (42)$$

Choosing the $Q$ matrix as:

$$Q = \begin{bmatrix} 9 & 0 & 0 \\ 0 & 0 & 0 \\ 0 & 0 & 0 \end{bmatrix}, \text{ and } R = 1,$$

$$J(x, t) = \int_t^{t_f} L(x^T \begin{bmatrix} 9 & 0 & 0 \\ 0 & 0 & 0 \\ 0 & 0 & 0 \end{bmatrix}, x + u^T u) \, d\tau$$

The command "lqr" in MATLAB may be used to determine the LQR problem's solution. Thus, the solution to the Riccati equation

$$SA^* + A^{*T}S + Q - SB^* R^{-1} B^{*T} S = 0$$

is determined as:

$$S = \begin{bmatrix} 4.8731 & 0.2831 & -0.3413 \\ 0.2831 & 0.0164 & -0.0198 \\ -0.3413 & -0.0198 & 0.0239 \end{bmatrix}$$

And,

$$K = [-4.0959 \quad -0.2380 \quad 0.2869]$$

The state feedback control is:

$$u^* = -R^{-1}B^{*T}S(t)x = -[-4.0959 \quad -0.2380 \quad 0.2869]$$
$$= [4.0959 \quad 0.2380 \quad -0.2869]\, x$$

The Simulink model of linearized and nonlinearized system optimal state feedback control is shown in Fig. 6 and Fig. 7, respectively.

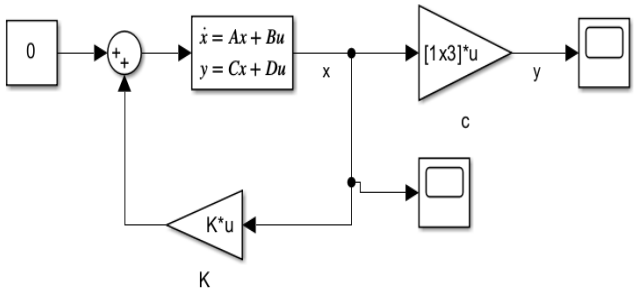

Fig. 6. Simulink model of linearized system optimal state feedback control

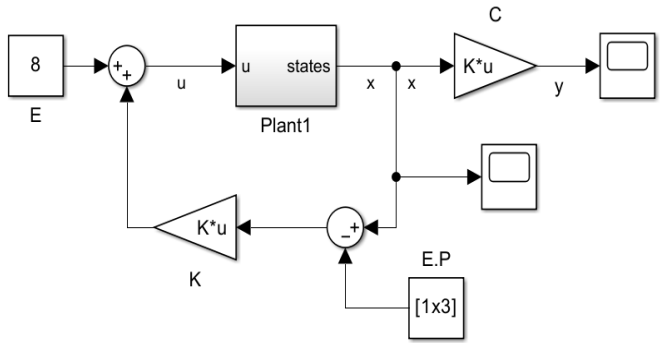

Fig. 7. Simulink model of a linearized system with optimal state feedback control

IV. SIMULATION RESULTS

The simulation results of the system for different control strategies are presented in the following subsections.

*a. State Feedback Control Simulation Results*

The simulation results of the MBSS control using a state feedback approach are presented in this subsection. The output waveform and the response of the states of the linearized system are presented in Fig. 8 and Fig. 9, respectively, whereas those of the nonlinear system are presented in Fig. 10 and Fig. 11, respectively. The systems are simulated for 50 s.

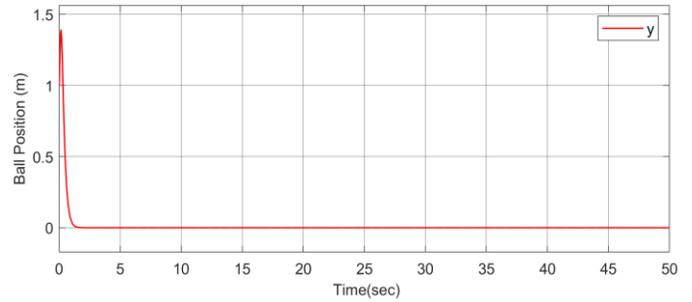

Fig. 8. Simulation result of the output response of the linearized state feedback control.

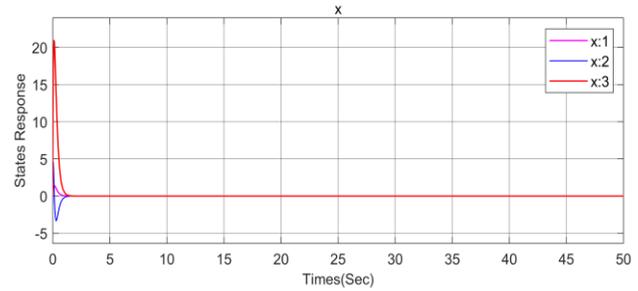

Fig. 9. Simulation result of the state response of the linearized state feedback control.

As shown in the simulation results of the linearized system (Fig. 8–9), the linearized system accurately stabilized the system to zero as expected after some swing. In contrast, the nonlinear system (Fig. 10–11) requires more time and exhibits several oscillations before reaching stability. These oscillations occur due to the greater control effort needed to drive the ball into its equilibrium point.

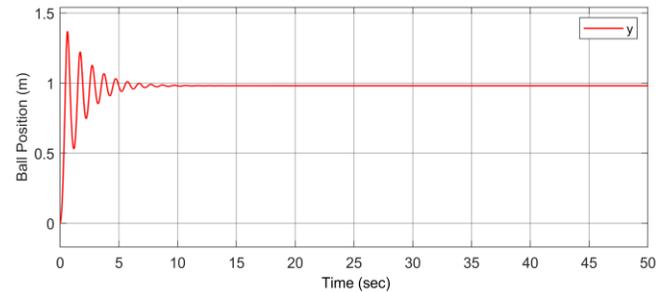

Fig. 10. Simulation result of the output response of the nonlinear state feedback control.

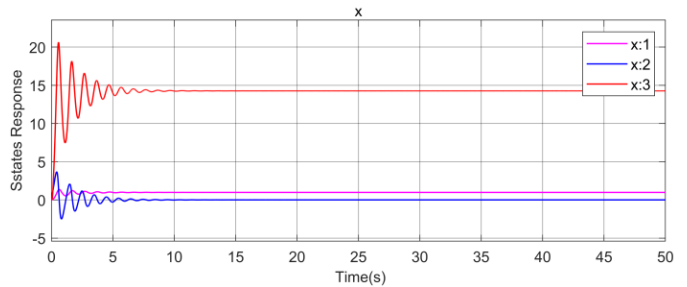

Fig. 11. Simulation result of the state response of the nonlinear state feedback control.

### b. Full-Order Observer Simulation Results

The simulation results of the MBSS control using an observer approach are presented in this subsection. The output waveform of the ball position, and the response of the states of the linearized system are presented in Fig. 12 and Fig. 13, respectively, whereas those of the nonlinear system are presented in Fig. 14 and Fig. 16, respectively. As shown in Fig. 12 – 13, the linearized observer system performed as expected, as it stabilized at the zero, after some swings.

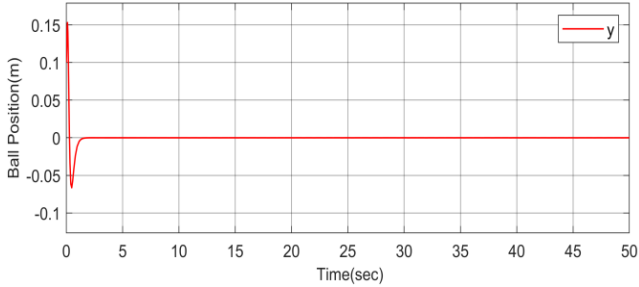

Fig. 12. Simulation result of the output response of the linear system observer.

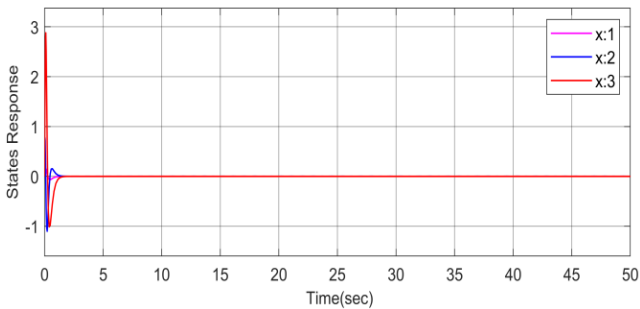

Fig. 13. Simulation result of the state response of the linear system observer.

On the other hand, as shown in Fig. 14 – 15, when the nominal value of 0.06 is used, the ball position based on the nonlinear system observer converged to the equilibrium point with a minor oscillation, compared to the performance of the system when only the state feedback control is used. However, when the nominal value of 0.8 is used, we observed an increased oscillation in the output waveforms of the system.

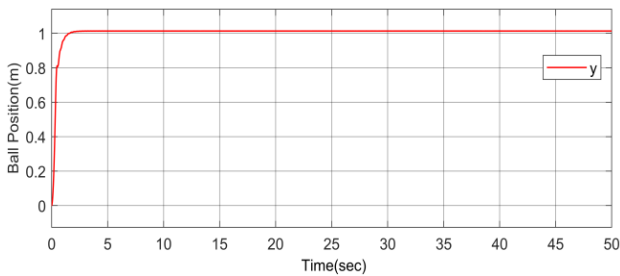

Fig. 14. Simulation result of the output response of the nonlinear system observer.

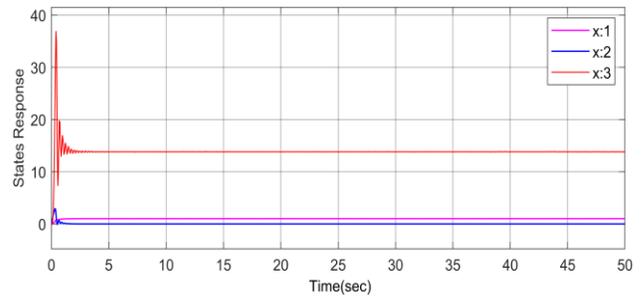

Fig. 15. Simulation result of the states' response of the nonlinear system observer.

### c. Optimal State Feedback Control Simulation Results

The simulation results of the MBSS control using an optimal state feedback control based on LQR are presented in this subsection. The output waveform of the ball position and the response of the states of the linearized system are presented in Fig. 16 - 17, whereas those of the nonlinear system are presented in Fig. 18 - 19.

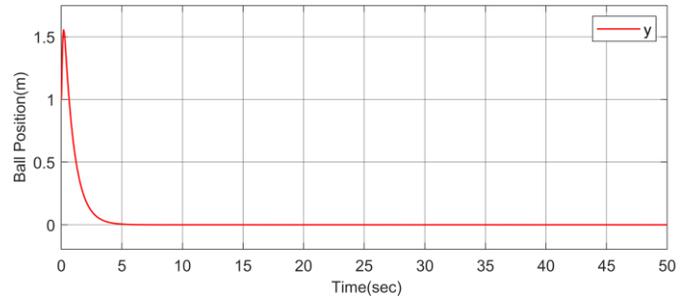

Fig. 16. Simulation result of the output response of the linearized optimal state feedback control.

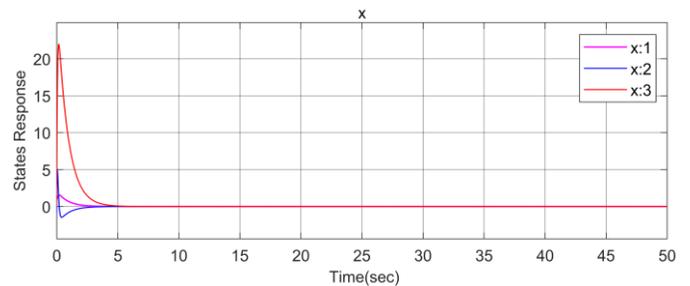

Fig. 17. Simulation result of the state response of the linearized optimal state feedback control.

As shown in the simulation results of the linearized system in Fig. 16–17, it can be seen that the linearized system accurately stabilized the system to zero as expected after some swing. However, this is not the case for the nonlinear system in Fig. 18 - 19, as it takes more time and several oscillations for the ball position to stabilize at a value close to 1 m. It can also be observed that the simulation results obtained from the state feedback approach presented earlier are similar to the optimal state feedback method.

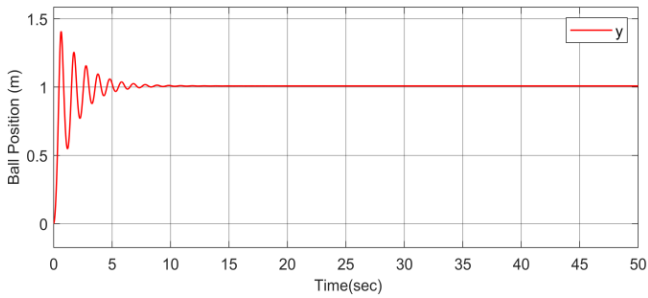

Fig. 18. Simulation result of the output response of the nonlinear optimal state feedback control.

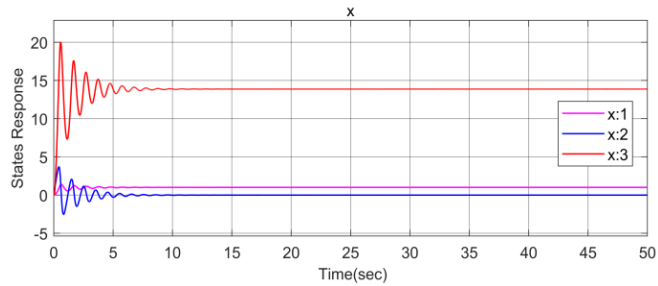

Fig. 19. Simulation result of the state response of the nonlinear optimal state feedback control.

## V. CONCLUSION

This study provides a detailed design of MBSS and analyzes its behavior in different operating states. The states and behavior of MBSS systems are always changing. Therefore, it is important to accurately determine its equilibrium points and robustly control the system output. This is achieved in this study by developing the state space model of the system and determining its equilibrium points using the feedback linearization

Following the above, we explored the observability and controllability status of the system, which is key in determining whether the poles of the system can be moved to any location with an input. The calculation results obtained showed that the system is controllable and observable. Then, we considered different approaches to controlling the MBSS, including the state feedback, full-order observer, and the optimal state feedback based on LQR, with the aim of controlling the ball's location to ensure that it performs satisfactorily. The simulation results showed that the linearized system will always perform satisfactorily irrespective of its initial conditions. However, due to the nonlinear nature of the MBSS, the system experienced some oscillations before it settled at the equilibrium. To achieve a more stable control of the system, it is recommended to use the smallest equilibrium point.

## REFERENCES

[1] M. Junaid Khan, D. Khan, S. Jabeen Siddiqi, S. Saleem, and I. Khan, "Design & Control of Magnetic Levitation System ED-4810: Review and Stability Test," *Univers. J. Electr. Electron. Eng.*, vol. 6, no. 4, pp. 191–202, Oct. 2019, doi: 10.13189/ujeee.2019.060402.
[2] L. A. Zadeh and World Scientific and Engineering Academy and Society, Eds., *Recent advances in signal processing, robotics and automation: proceedings of the 9th WSEAS International Conference on Signal Processing, Robotics and Automation (ISPRA '10) ; University of Cambridge, UK, February 20 - 22, 2010*. in Mathematics and computers in science and engineering. S.l.: WSEAS Press, 2010.
[3] J.-J. E. Slotine and W. Li, *Applied nonlinear control*. Englewood Cliffs, N.J: Prentice Hall, 1991.
[4] B.P._Lathi, "Signal_Processing_and_Linear_Systems" (b-ok.org).pdf."
[5] "Lin - 2007 - Robust control design an optimal control approach.pdf."
[6] M. F. Golnaraghi, B. C. Kuo, and M. F. Golnaraghi, *Automatic control systems*, 9th ed. Hoboken, NJ: Wiley, 2010.
[7] F. Lin, *Robust control design: an optimal control approach*. in RSP series in control theory and applications. Chichester: Wiley, 2007.